\documentclass[twocolumn,aps,prl,floats,showpacs]{revtex4}
\newcommand{\beq}{\begin{eqnarray}}
\newcommand{\eeq}{\end{eqnarray}}

\usepackage{epsfig}% Include figure files
\usepackage{dcolumn}% Align table columns on decimal point
\usepackage{bm}% bold math

\begin{document}
\bibliographystyle{prsty}
\title{Coulomb Blockade with Dispersive Interfaces}
\draft
\author{ A. Kamenev$^{1}$ and  A. I. Larkin$^{1,2}$ }

\address{$^1$ Department of Physics, University of Minnesota,
Minneapolis, MN 55455, USA}
\address{$^2$L. D. Landau Institute for Theoretical Physics,
Moscow 117940, Russia}

\date{\today}

%\maketitle

%\parbox{14cm}
\begin{abstract}
    {\rm What quantity controls the Coulomb blockade oscillations if
    the dot--lead conductance is essentially frequency--dependent ? We argue
    that it is the {\em ac dissipative}  conductance at the frequency
    given by the effective charging energy. The latter may be
    very different from the bare charging energy due to
    the interface--induced capacitance (or inductance). These observations are
    supported by a number of examples, considered from  the weak and strong
    coupling (perturbation theory vs. instanton calculus) perspectives.  }
\end{abstract}

 \maketitle

%\bcols
%\maketitle

%\begin{multicols}{2}

\bigskip

%\section{Introduction}

The quantum fluctuations of charge in a  Coulomb blockade (CB)
quantum dot  \cite{Likharev,Schon90} have been a subject of
intense studies (see Refs.~\cite{Aleiner,Grabert02} for a recent
review). It is widely accepted that the dimensionless (in units of
$G_Q=e^2/(2\pi \hbar)$) conductance, $G$,  of the interface
between the dot and a lead  controls the strength of the quantum
fluctuations. If $G<1$ (weak coupling) the fluctuations are  small
(possibly apart from the degeneracy points) and usually may be
accounted for by the perturbation theory in $G$
\cite{Glazman90,Grabert94,Grabert02}. On the other hand, for $G>1$
(strong coupling) the CB is basically destroyed by the
fluctuations. The remained weak  CB oscillations  may be described
in the instanton approach
\cite{Zaikin91,Grabert96,Nazarov99,Kamenev00,Andreev01,Feigelman02}.

What happens if the conductance of the interface is essentially
frequency--dependent, $G=G(\omega)$ ? Is it the dc conductance,
$G(0)$, or rather an ac one, $G(\Omega_c)$, (or may be none of
them) that governs the strength of the quantum fluctuations ? Is
there a single parameter that divides the regions of applicability
of the perturbation theory and the instanton calculations ?  What
energy scale controls the thermal fluctuations ? These are the
questions we address in the present paper \cite{foot6}.

We argue that it is the dissipative {\em ac} conductance, $\Re e\,
G(\Omega_c)$, that controls the quantum fluctuations and
seamlessly divides between the two methods of description. The
relevant frequency $\Omega_c$ has the meaning of an effective
charging energy of the dot, $\Omega_c \sim e^2/(C+\tilde C)$,
where $C$ is the bare capacitance of the dot and $\tilde
C(\omega)$ is the parallel capacitance induced by the dispersive
interface. Frequency $\Omega_c$ is the solution of the following
self--consistent equation:
\begin{equation}
G_Q^{-1}\left( C +\tilde C(\Omega_c)\right) =\Omega_c^{-1} \,;
\hskip .4cm
 \tilde C(\omega) \equiv G_Q \,\partial_\omega G(i\omega)\, .
                                     \label{eq1}
\end{equation}
The effective charging energy, $\Omega_c$, serves also as the
characteristic scale for the  thermal smearing of the CB. The
induced capacitance, $\tilde C(\omega)$, is not necessarily
positive and therefore $\Omega_c$ may be on the either side of the
bare charging energy, $E_c=e^2/(2C)$.

To illustrate these points we consider three specific examples. In
one of them (tunnelling barrier): $G(0) \approx G(\Omega_c)$. In
the second (resonant impurity interface) the dc conductance is
parametrically larger than the ac one, $G(0) \gg G(\Omega_c)$.
Finally, in the third example (shallow 2d lead) the situation may
be  reversed: $G(0)\ll G(\Omega_c)$.  We treat all the examples
from both  weak and  strong coupling perspectives. While the
former is relatively straightforward, the latter requires certain
technical adjustments to the way the instanton calculus is usually
applied to the CB problem.

The CB is described by the following partition function
\begin{equation}
Z(q) = \sum\limits_{W=-\infty}^\infty e^{i2\pi q W}
\!\!\!\!\!\!\!\! \int\limits_{\phi(\beta)-\phi(0) =2\pi
W}\!\!\!\!\!\!\!\! {\cal D}\phi\, \, e^{-S[\phi]}\,\, ,
                                     \label{partition}
\end{equation}
where $q$ is the external charge controlled by the gate voltage
and $W$ are integer winding numbers \cite{Schon90}. The imaginary
time phase action has the form \cite{foot2}
\begin{equation}
S[\phi]= \int\limits_0^\beta \!\! d\tau\, \frac{\dot
\phi^2(\tau)}{4E_C}\, - \,
\int\!\!\!\!\int\limits_{\!\!\!\!0}^\beta \!\! d\tau d\tau'\,
 e^{i\phi(\tau)} K(\tau-\tau') e^{-i\phi(\tau')} \, .
                                          \label{action}
\end{equation}
The boson Matsubara transform $K(\omega_m)=K_m$ of the kernel
$K(\tau)$ is given by
\begin{equation}
K_m = \mbox{Tr}{1\over \beta} \sum\limits_{\epsilon_n} \hat {\cal
W}_n {\cal G}_n^{(l)}\hat {\cal W}_n^\dagger \left\{ {\cal
G}_{n+m}^{(d)} - {\cal G}_{n}^{(d)} \right\} \, ,
                                            \label{cond}
\end{equation}
where $\hat {\cal W}_n=\hat {\cal W}(\epsilon_n)$ is the
tunnelling matrix connecting the dot and the lead and  ${\cal
G}^{({d,l})}_n = {\cal G}^{({d,l})}(\epsilon_n)$ are the fermion
Green functions of the dot and lead correspondingly. The linear ac
conductance of the interface is determined by  the kernel, $K_m$.
Indeed, employing the Kubo formula, one finds for  the Matsubara
transform of the conductance $G_m= -4\pi |\omega_m|^{-1}\Re e\,
K_m$. After analytical  continuation from the upper complex
half--plane, $\omega_m\to -i\omega$, one obtains the complex ac
interface conductance $G(\omega)$. The kernel $K_m$ is also
related to the non--linear current--voltage characteristic of the
interface \cite{Mahan}. Namely, $I(V) = 2e\, \Im m
\{K\left|_{\omega_m \to ieV}\right.\}$ is the current through the
interface, provided that the voltage $V$ is applied across it. We
shall mostly employ the ac interpretation, although the $I-V$
language is also possible.

In the weak coupling regime  one  expands $\exp\{-S[\phi]\}$ in
the power series over the $K$--term in Eq.~(\ref{action}).
Performing then the Gaussian $\phi$--integration and
$W$--summation, one finds for the average number of electrons on
the dot, $\delta N(q) \equiv q + (2\beta E_c)^{-1} \partial \ln
Z/\partial q\, $:
\begin{equation}
\delta N(q) =  \int\limits_{-\infty}^{\infty}\!\! d \tau\, \tau
K(\tau)\,\,  e^{-E_C(|\tau|-2q\tau)}  \, .
                                      \label{weak}
\end{equation}
We have restricted ourselves to the temperature range $\beta E_C >
1$ and thus extended the integration to infinity. For a closed dot
the electron number stays constant for $|q|<1/2$ and jumps by one
at the degeneracy points, $q=\pm 1/2$. A  weak coupling,
$K(\tau)$, makes the electron number to change smoothly in between
the degeneracy points. It is clear from Eq.~(\ref{weak}) that,
apart from the immediate vicinity of $q=\pm 1/2$ \cite{Matveev91},
it is the short time, $|\tau| < E_c^{-1}$, behavior of $K(\tau)$
that determines $\delta N(q)$. One expects, thus, Eq.~(\ref{weak})
to be valid if $\Omega_c\approx E_c$ and $\Re e\, G(\Omega_c) \ll
1$.

If  this is not the case one may employ the instanton approach. To
this end one has to find  stationary configurations of  the action
(\ref{action}) with a fixed winding number, e.g. $W= 1$. We are
unable to execute this program for an arbitrary $K(\tau)$ and
employ instead the variational method. As a trial variational
solution we take a family of Korshunov instantons
\cite{Korshunov}:
\begin{equation}
e^{i\phi(\tau)} =  \frac{e^{2\pi i\tau/\beta} -z}{1- e^{2\pi
i\tau/\beta} z^*}  \, \, ,
                                     \label{korshunov}
\end{equation}
where $z$ is a complex number with $|z| <1$  that determines
location and the width of the instanton. The action of a trial
configuration is, in general, complex. Its imaginary part is
simply a shift  of the background charge, $q$, which affects only
a phase of the CB oscillations. Hereafter we focus on the real
part. Substituting the trial phase configuration,
Eq.~(\ref{korshunov}), into Eq.~(\ref{action}), one finds
\begin{equation}
\Re e\,S(\Omega) = {\pi \Omega\over 2 E_c} +
\frac{\pi\partial}{\partial(\beta\Omega)}\,
\sum\limits_{m=0}^{\infty} \left( 1- {2\pi\over\beta \Omega}
\right)^{\!\!m}\!\! G(\omega_m)\, ,
                                            \label{variational}
\end{equation}
where $\beta \Omega/(2\pi) \equiv (1-|z|^2)^{-1}\geq 1$ is the
inverse width of the instanton. One should minimize now $Z_1\sim
(\beta\Omega)\exp\{-\Re e\, S(\Omega)\}$ with respect to the
instanton width, $\Omega^{-1}$. The factor $\beta\Omega$
originates from the integration over the zero mode associated with
the instanton location,  $\arg z$. The minimum is reached at
$\Omega=\Omega_c$ that is a solution of the stationary point
equation  $d\, \Re e\, S(\Omega)/d\Omega = \Omega^{-1}$. At low
temperature this equation essentially coincides with
Eq.~(\ref{eq1}). The optimal instantons are usually narrow,
$\beta\Omega_c \gg 1$, that justifies the non--interacting
instanton gas approximation. Summing up the gas of optimal
instantons and anti--instantons \cite{Grabert96,Feigelman02}, one
finds for the average number of electrons
\begin{equation}
\delta N(q) = q - c\, \frac{\Omega_c}{E_c}\,\, e^{-\Re e\,
S(\Omega_c)}\, \sin(2\pi q + \varphi) \, ;
                                      \label{strong}
\end{equation}
with a model dependent proportionality coefficient, $c$, and
phase, $\varphi$. This result is applicable if $\Re e\,
S(\Omega_c) \gg 1$.

We have also employed the trial  configuration of the form
$\phi(\tau)=4\arctan\left(\exp\{\Omega\tau\}\right)$, that  is
known to be a solution for the Josephson tunnelling problem
\cite{Efetov80}. The results differ from those obtained with
Eq.~(\ref{korshunov}) only by numerical coefficients. That makes
us believe that our variational approach is  parametrically
accurate. We turn now to the applications of these results for the
specific examples.

(i) {\em  Frequency--independent} conductance: $G_m = G_0$. This
is the case e.g. if both lead and dot have continuous spectrum,
${\cal G}^{({d,l})}_n = i\pi\nu^{({d,l})}\mbox{sign}(\epsilon_n)$,
and energy independent tunnelling matrix elements, ${\cal W}$. In
the low temperature limit the  kernel has the form $K(\tau) =
G_0/(2\pi\tau)^2$. Employing Eq.~(\ref{weak}), one finds for a
weakly coupled dot \cite{Glazman90} $\delta N(q)={G_0\over
(2\pi)^2}\ln {1+2q \over 1-2q}\,$. In the strong coupling regime
the trial phase, Eq.~(\ref{korshunov}), is the exact solution of
the saddle point equation. For $\Omega\ll E_c$ the action $\Re e\,
S(\Omega) = G_0/2$ is $\Omega$--independent. As a result, there is
an additional zero mode associated with the instanton width.  The
presence of such zero--mode  complicates  calculations of the
pre--exponential factor. We quote here only the result
\cite{Grabert96,Andreev01,Feigelman02}: $c\, \Omega_c$ in
Eq.~(\ref{strong}) must be substituted by $(2\pi)^{-1}E_c G_0^2
\ln(\beta E_c)$. The additional zero--mode leads to the
instanton--instanton interactions that break  the non--interacting
instanton gas approximation at  temperature $\beta^{-1}\approx E_c
G_0^2\exp\{-G_0/2\}$. At smaller temperature the amplitudes of the
higher $q$--harmonics are probably of the same order as the first
one. That may lead to a nonanalytic behavior at the degeneracy
points, $q=\pm 1/2$, in the zero temperature limit, similar to
that of Ref.~\cite{Matveev95}. As expected, the unit conductance,
$G_0\approx 1$, separates the strong and weak coupling regimes.

(ii) {\em  Resonant impurity. } Consider an interface where the
particle exchange between the dot and  bulk   takes place only
through resonant impurity levels. For a single channel   this
model was solved in Ref.~\cite{Matveev01}. We shall assume for
simplicity that all the levels are at the  Fermi energy and have
the same coupling to the bulk (width), $\Gamma\ll E_c$. The Green
function of the lead is given by that of the resonant impurity
level: ${\cal G}_n^{(l)} =
(i\epsilon_n+i\Gamma\mbox{sign}(\epsilon_n))^{-1}$. With the help
of Eq.~(\ref{cond}), one finds for the kernel:
\begin{equation}
K_m=-\frac{G_0 \Gamma}{4\pi}\, \ln\left(
1+\frac{|\omega_m|}{\Gamma} \right)\, .
                                                           \label{Kres}
\end{equation}
Accordingly the linear conductance takes the form $G(\omega) =
G_0\Gamma i\omega^{-1} \ln(1- i \omega/\Gamma)$, where $G_0$ is
the dc resonant conductance proportional to the number of
impurities and their coupling to the dot \cite{foot7}. The short
time, $\Gamma\tau\ll 1$, limit of the kernel is  $K(\tau)=
G_0\Gamma/ (8\pi|\tau|)$. Employing Eq.~(\ref{weak}), one obtains
for $\delta N(q)$ not too close to the degeneracy points, $q=\pm
1/2$ \cite{Matveev01}:
\begin{equation}
\delta N(q)=\frac{G_0\Gamma}{8\pi E_c}\left(\frac{1}{1-2q} -
\frac{1}{1+2q}\right)\, .
                                                     \label{weakres}
\end{equation}
Notice, that the criterion of the weak coupling is $G_0\Gamma/E_c
\ll 1$ \cite{Matveev01}, while the dc conductance, $G_0$, may be
large. In such a case  Eq.~(\ref{eq1}) is solved by
$\Omega_c\approx E_c$. Since  $\Re e \, G(\omega) = \pi
G_0\Gamma/(2\omega)$ for $\omega\gg \Gamma$, one finds that
$\delta N(q)\sim \Re e\, G(\Omega_c)$. As a result, the weak
coupling approximation is valid as long as $\Re e\, G(\Omega_c)
\ll 1$.

In the strong coupling regime, employing Eq.~(\ref{variational}),
one finds for the variational action
\begin{equation}
\Re e\,S(\Omega) = {\pi \Omega\over 2 E_c} +  \frac{
G_0\Gamma}{2\Omega}\, e^{\Gamma/\Omega}
E_1\left(\Gamma/\Omega\right)
 \,  ,
                                            \label{resvariational}
\end{equation}
where $E_1$ is the exponential  integral. This action possess  two
stationary points as a  function of the instanton width,
$\Omega^{-1}$. One is the stable minimum  at $\Omega_c =
\sqrt{G_0\Gamma E_c/(2\pi)}\, \ln^{1/2}(G_0 E_c/\Gamma)$, with the
action $\Re e\,S(\Omega_c)= \pi \Omega_c/E_c$. The other is the
unstable plateau  at $\Gamma\Omega^{-1} > 1$ with the action given
by  half of the  dc conductance: $\Re e\,S(0)=G_0/2$, in agreement
with (i). One may check that $\Omega_c$ is indeed the solution of
Eq.~(\ref{eq1}). Up to logarithmical factor \cite{foot5}, $\Re
e\,S(\Omega_c)\approx \Re e\, G(\Omega_c)$ and the condition $\Re
e\, G(\Omega_c) \approx\sqrt{G_0\Gamma/E_c}\gg 1$ justifies
validity of the instanton approach.

At not very small temperature, the stable minima at $\Omega_c$
(narrow instanton) is the only relevant one and the electron
number is given by Eq.~(\ref{strong}). The contribution of the
wide instantons with $\Omega <|\tilde \Gamma|$  is exponentially
smaller $\sim G(0)^2\ln (\beta|\tilde \Gamma|)\, e^{-G(0)/2}$ and
may be disregarded as long as $G(0) \gg \Re e\, G(\Omega_c)
\approx \Re e\, S(\Omega_c)$. Nevertheless, due to the phenomena
mentioned in (i), the wide instantons may prove to be important in
an immediate vicinity of the degeneracy points and at very small
temperature, $\beta^{-1}< E_c G_0^2\exp\{-G(0)/2\}$. If so, then
$G(0)$, rather than $\Re e\, G(\Omega_c)$, controls the CB in that
narrow range.

On the high temperature side the CB oscillations are washed out at
$T > \Omega_c$. Notice, that in the strong coupling regime
$\Omega_c
>E_c$ and therefore  the (weak) CB, Eq.~(\ref{strong}),  is more
``temperature--tolerant'' than in the weak coupling. The reason is
that the induced capacitance, $\tilde C$, is negative (it is
actually an inductance) and thus the effective charging energy is
increased.

(iii) {\em Shallow 2d lead}. Consider a quantum dot coupled
through a tunnelling barrier to a shallow 2d gas. The bottom of
the 2d conductance band is at energy $\varepsilon_0$ below
($\varepsilon_0<0$) or above ($\varepsilon_0>0$) of the dot's
Fermi energy. In the former case the dc conductance between the
dot and 2d gas is given by the product of their densities of
states and the tunnelling matrix elements and is denoted by $G_0$.
In the latter case the dc conductance is zero. We shall examine
the effect of the proximity  to such a lead ($\varepsilon_0\ll
E_c$ hereafter) on the CB in the dot.

The straightforward calculation yields for the coupling kernel:
%\begin{equation}
%{\cal G}^{(l)}(\tau) = {\pi \nu^{(l)} \over \tau} \left\{
%\begin{array}{ll}
%1-\theta(-\tau)\, e^{-\varepsilon_0\tau}\, , & \,\,
%\varepsilon_0<0\, ;
%\\
%\theta(\tau)\, e^{-\varepsilon_0\tau}\,, & \,\, \varepsilon_0>0\,
%.
%\end{array}
%\right.
%\end{equation}
%As a result, one finds for the kernel
\begin{equation}
K(\tau) = {G_0 \over (2\pi\tau)^2} \left\{
\begin{array}{ll}
1-\theta(-\tau)\, e^{-\varepsilon_0\tau}\, , & \,\,
\varepsilon_0<0\, ;
\\
\theta(\tau)\, e^{-\varepsilon_0\tau}\,, & \,\, \varepsilon_0>0\,
.
\end{array}
\right.
                                 \label{Kshallow}
\end{equation}
The corresponding conductance, $G(\omega)$, obeys the symmetry
relation $G_-(\omega)= G_0 - G_+(\omega)$, where the subscripts
$\pm$ denote the sign of $\varepsilon_0$. In the small temperature
limit, $\beta\varepsilon_0\gg 1$ one finds for the dissipative
conductance
\begin{equation}
\Re e\, G_+(\omega)={G_0 \over 2}\,\,  \theta (|\omega|-
\varepsilon_0) \left(1-{\varepsilon_0\over |\omega|}\right)\, .
                                            \label{realG}
\end{equation}
Notice, that $\Re e\, G(\omega) =0$ for $|\omega| \leq
\varepsilon_0$ if the 2d band is empty, $\varepsilon_0 >0$. On the
other hand, in the high frequency limit, $\omega\gg
|\varepsilon_0|$, $\Re e\, G(\omega) =G_0/2$ irrespective to the
sign of $\varepsilon_0$. For the reactive conductance one obtains
$\Im m\, G_+(\omega) = G_0/(2\pi) f(|\omega|/\varepsilon_0)$,
where we denoted  $f(x) = \ln |x-1|-\ln|x+1| - x^{-1} \ln|x^2-1|$.

In the weak coupling limit, employing Eqs.~(\ref{weak}) and
(\ref{Kshallow}), one finds for the average number of electrons:
\begin{equation}
\delta N(q) = {G_0 \over (2\pi)^2} \left\{
\begin{array}{ll}
\ln{1+2q\over 1-2q} - \ln(1+ |\varepsilon_0|/E_c+2q)\, , & \,
\varepsilon_0<0\, ;
\\
- \ln(1+ \varepsilon_0/E_c-2q)\,, & \,\, \varepsilon_0>0\,
\end{array}
\right.
                                         \label{shalloweak}
\end{equation}
(we have subtracted an irrelevant constant $-(G_0 /(4\pi)^2)\ln
\Lambda/E_c$, where $\Lambda$ is the 2d band--width). For the
shallow filled 2d gas ($\varepsilon_0<0$) the result is only
slightly different from (i). Surprisingly, for the empty 2d band
($\varepsilon_0>0$) one finds a $q$--dependence of $\delta N$,
despite of the fact that the dc conductance is strictly zero. This
dependence is parametrically similar to that of the filled 2d gas,
however does not exhibit singularities, indeed $|q| <1/2$. It is
associated with the virtual transitions between the dot and the
empty 2d band.  Since $\varepsilon_0 \ll E_c$, such transitions
are possible and the coupling strength is determined by $G_0$
only. For $G_0 < 1$ Eq.~(\ref{eq1}) is solved by $\Omega_c\approx
E_c$. The applicability of the weak coupling approximation is
therefore controlled by the condition $\Re e\, G(\Omega_c) = G_0/2
\ll 1$ (and {\em not} $G(0) \ll 1$ !).

We shall   approach now the problem from the strong coupling
perspective. To this end we employ the variational expression,
Eq.~(\ref{variational}). For $\varepsilon_0<0$ the variational
action is $\Re e\,S= G_0(1 -\Omega/(\pi\varepsilon_0))/2$ for
$\Omega\ll |\varepsilon_0|$, while it approaches   $\Re e\,
S=G_0/4 +\pi\Omega/E_c$ for $\Omega\gg |\varepsilon_0|$. There is
a broad minimum at $|\varepsilon_0| \ll \Omega_c <E_c$. As a
result, $\Re e\, S(\Omega_c) = G_0/4$  and the electron number is
given by Eq.~(\ref{strong}) (possibly apart from the immediate
vicinity of $q=\pm 1/2$). This result is valid for $\Re e\,
G(\Omega_c) = G_0/2 \gg 1$.

For the empty 2d band, $\varepsilon_0 >0$, the variational action
is  $\Re e\,  S = G_0\Omega/(2\pi\varepsilon_0) $  for $\Omega \ll
\varepsilon_0$ and riches $G_0/4$ at $\varepsilon_0\ll \Omega <E_c
$. Naively, it has a minimum at $\Omega_c \to 0$ with $\Re e\,
S(0)=0$. One has to include, however, the fluctuation part into
the stationary point  equation, as discussed after
Eq.~(\ref{strong}). As a result, for $G_0 \gg 1$  one finds the
stable extremum at  $\Omega_c = 2\pi \varepsilon_0/G_0$. The
corresponding electron number is $\delta N(q) = q -
c\varepsilon_0/(G_0 E_c)  \sin (2\pi q)$.  Despite being obtained
in the instanton approach, the correction is not exponentially
small. The formal reason is that $\Re e\,S(\Omega_c) = 1$, that
makes the above  result an order of magnitude estimate, at best.
Fortunately, one can solve the problem in a different way.

As was shown above the ''straight'' instantons, $\phi_W(\tau) =
2\pi W\tau/\beta$, are the most stable. One can, thus, calculate
the partition function as a sum over such instantons. The action
is the Matsubara transform of the kernel: $\Re e\, S[\phi_W] =
-\beta \Re e\, K(\omega_W)\approx W^2 G_0/(2\beta\varepsilon_0)$.
Employing Eq.~(\ref{partition}), one finds $\ln Z(q) = - \beta q^2
(2\pi^2\varepsilon_0/G_0)$; valid at small temperature,
$\beta^{-1} \ll \varepsilon_0/G_0$. As a result,
\begin{equation}
\delta N(q) = q \left( 1- \frac{2\pi^2\varepsilon_0}{G_0 E_c}
\right)\, ; \hskip 1cm  |q|<1/2\,
                                                 \label{shallowstrong}
\end{equation}
and there are small jumps by $\pi\varepsilon_0/(G_0 E_c) \ll 1$ at
the degeneracy points $q=\pm 1/2$. The presence of such jumps
could be anticipated from the absence of singularities in the weak
coupling expression, Eq.~(\ref{shalloweak}).

Equation~(\ref{shallowstrong}) states that there is a perfect CB
with the reduced (screened) charging energy $\pi\Omega_c=2\pi^2
\varepsilon_0/G_0$. Indeed, according to Eq.~(\ref{eq1}) the
induced  capacitance is given by $\tilde C(\omega) = e^2
G_0\varepsilon_0/(2\pi\omega)^2 \ln(1+(\omega/\varepsilon_0)^2)$.
As a result, one finds $\tilde C= e^2 G_0/(4\pi^2\varepsilon_0)
\gg C$ at small frequency, $\omega \ll \varepsilon_0$. The bare
capacitance, $C$, appears to be completely screened and
substituted by $\tilde C$. Solving  Eq.~(\ref{eq1}) with such
$\tilde C$, one obtains $\Omega_c= 2\pi \varepsilon_0/G_0$ that
justifies the  small frequency assumption, provided $G_0 \gg 1$.
The amazing twist is that $\Re e\, G(\Omega_c) =0$ and quantum
fluctuations are practically absent. Therefore the larger $G_0$ is
-- the {\em weaker} is the coupling to the lead ! By increasing
$G_0$ one can restore the perfect CB with a very small charging
energy, however. The (exponential) temperature smearing sets in at
the temperature $\beta^{-1} \approx \Omega_c \ll E_c$, limiting an
indefinite  increase of $G_0$.

Strictly speaking, the system is always in the weak coupling
regime, $\Re e\, G(\Omega_c)\ll 1$ (possibly apart from
$G_0\approx 1$).  The perturbative result, Eq.~(\ref{shalloweak}),
is valid, though, only for $G_0\ll 1$. For $G_0\gg 1$
Eq.~(\ref{shallowstrong}) is to be employed. (It is in parametric
agreement  with the  variational estimate, although the
$q$--dependence of the latter is actually wrong.)  The two results
match at  $G_0\approx 1$. Indeed, consider e.g. $\chi= \partial
\delta N(q)/\partial q|_{q=0}$. For $G_0 \ll 1$ one finds $\chi =
(2\pi^2)^{-1} G_0/(1+\varepsilon_0/E_c)$, while for $G_0 \gg 1$
$\chi = 1 - 2\pi^2\, \varepsilon_0/(G_0 E_c)$. Since
$\varepsilon_0\ll E_c$, these two (weak coupling) expressions
perfectly match at $G_0 = 2\pi^2$.

To summarize: we have argued that not too close to the degeneracy
points the quantum fluctuations of charge are governed by the
dissipative ac conductance $\Re e\, G(\Omega_c)$. The effective
charging energy, $\Omega_c$, has to be  found as the solution of
Eq.~(\ref{eq1}). It may substantially deviate from the bare
charging energy, $E_c$, in either direction. For $\Re e\,
G(\Omega_c) \ll 1$  and low enough temperature, $\beta\Omega_c\gg
1$, the charge on the dot is approximately quantized. Formally it
stems from the fact that the large  winding numbers, $W \approx
\beta\Omega_c\gg 1$, give a dominant contribution to the partition
function. In the opposite limit, $\Re e\, G(\Omega_c) \gg 1$, only
the instantons with the smallest $W$'s are important, while
$\Omega_c$ acquires the meaning of the optimum inverse instanton
width. In both cases for $\beta^{-1} > \Omega_c$ the thermal
fluctuations smear the CB. On the other hand, very close to the
degeneracy points and at small temperature, the relevant parameter
may prove to be the dc conductance, $G(0)$.  A detailed study of
the last issue is beyond the scope of the present letter.

We are grateful to L.~I.~Glazman and M.~Pustilnik  for numerous
useful discussions. AK was partially supported  by the BSF grant N
9800338. AIL was partially supported by the NSF grant N 0120702.

%\end{document}

%\ecols

%\end{multicols}


\begin{thebibliography}{100}
\bibitem{foot7} The dc conductance $G_0 = N\Gamma_d/\Gamma$, where
$N$ is the number of resonant levels and $\Gamma_d$ is the
coupling between the dot and an impurity. We assume $N>1$ and
$\Gamma_d <\Gamma$ \cite{foot2}.
\bibitem{foot5} It is  actually $G(i\,\Omega_c)$, rather than
$\Re e\, G(\Omega_c)$ that controls the quantum fluctuations. In
all examples, we have been focusing on, they happen to be almost
the same.
\end{thebibliography}
\end{document}